\documentclass[english,11pt,a4paper]{article}
\pdfoutput=1 

\usepackage{jheppub} 
\usepackage{graphicx}
\usepackage{enumerate}
\usepackage{bm}
\usepackage{slashed}
\usepackage{xcolor}

\usepackage{amssymb}
\usepackage{amsmath}
\usepackage{epsfig}
\newcommand{\be}{\begin{equation}}
\newcommand{\ee}{\end{equation}}
\newcommand\beq{\begin{eqnarray}}
\newcommand\eeq{\end{eqnarray}}

\newcommand{\CO}{{\cal O}}
\newcommand{\CM}{{\cal M}}

\newcommand{\nn}{\nonumber}

\newcommand{\mybar}[1]%
        {\kern 0.6pt\overline{\kern -0.6pt#1\kern -0.6pt}\kern 0.6pt}
\begin{document}

\title{Chiral Anomalous Dispersion}

\author[a, b]{Andrey Sadofyev}
\author[c, d]{,$~~$Srimoyee Sen}

\affiliation[a]{Theoretical Division, MS B283, Los Alamos National Laboratory, Los Alamos, NM 87545}
\affiliation[b]{Center for Theoretical Physics, Massachusetts Institute of Technology, Cambridge, MA 02139}
\affiliation[c]{Institute for Nuclear Theory, Seattle, WA 98105}
\affiliation[d]{Department of Physics, University of Arizona, Tucson, AZ 85721}

\emailAdd{sadofyev@lanl.gov}
\emailAdd{srimoyee08@gmail.com}

\preprint{\begin{minipage}[t]{0.2\textwidth}\raggedleft\footnotesize{INT-PUB-17-052}\end{minipage}}
 
\abstract{ 
The linearized Einstein equation describing graviton propagation through a chiral medium appears to be helicity dependent. We analyze features of the corresponding spectrum in a collision-less regime above a flat background. In the long wave-length limit, circularly polarized metric perturbations travel with a helicity dependent group velocity that can turn negative giving rise to a new type of an anomalous dispersion. We further show that this chiral anomalous dispersion is a general feature of polarized modes propagating through chiral plasmas extending our result to the electromagnetic sector.
}

\maketitle
\flushbottom

\section{Introduction}
It is well known by now that chiral media can exhibit a novel class of transport phenomena that are anomalous in nature. In a plasma of massless fermions with interactions that preserve chiral symmetry, the axial anomaly results in a modification of electric, axial, and energy-momentum transports, for a review see \cite{Kharzeev:2015znc}. Usually chiral effects are distinguished by a source to be vortical or magnetic. The most widely discussed examples of the anomalous transport are contributions to the electric current 
\beq
J^i=\sigma_{CME}B^i~~,~~J^i=\sigma_{CVE}\Omega^i
\eeq
which are referred to as chiral magnetic effect (CME) and chiral vortical effect (CVE). These phenomena may considerably modify the medium evolution and are suggested to play an essential role in various systems such as quark-gluon plasma, Dirac and Weyl semi-metals, and primordial plasma (see e.g. \cite{Joyce:1997uy, Li:2014bha, Kharzeev:2015znc}).

A particularly violent modification of the medium dynamics by chiral effects is tied with a set of instabilities 
\cite{Joyce:1997uy, Akamatsu:2013pjd, Khaidukov:2013sja, Kirilin:2013fqa, Tuchin:2014iua, Avdoshkin:2014gpa, Buividovich:2015jfa, Kaplan:2016drz, Hattori:2017usa}. For instance, if the anomalous back-reaction of the medium is taken into account, Maxwell equations support an exponentially growing mode of a helical magnetic field \cite{Joyce:1997uy, Akamatsu:2013pjd}. This process transfers microscopic chirality of fermions to macroscopic helicity of magnetic fields and originates from the mixing of these two quantities. Such transfer is a general ingredient among all chiral instabilities \cite{Avdoshkin:2014gpa, Yamamoto:2015gzz, Zakharov:2016lhp, Kirilin:2017tdh}.

Not unexpectedly, there are examples of chiral effects sourced by an external gravitational field\footnote{Note that the regular chiral effects may also gain corrections due to the presence of a gravitational field, see e.g. \cite{Basar:2013qia, Flachi:2017vlp}.}. Indeed, already the derivation of CVE requires one to consider two point function of the current and the stress energy tensor \cite{Landsteiner:2011cp}. This mixing between responses to the medium velocity and the $0i$ component of the metric perturbation is known for a long time. Further analysis shows that a transverse traceless metric perturbation also leads to new contributions to the stress-energy transport \cite{Jensen:2012kj, Manes:2012hf}. As a simple example, one may consider an anomalous response caused by a gravitational wave (GW) propagating through the medium. Without loss of generality we choose the GW momentum to be in the $z$-direction denoting it as $q_3$ and its absolute value as $q$, then the only non-zero components of a GW field $h^{\mu\nu}=g^{\mu\nu}-\eta^{\mu\nu}$ are $h^{11}=-h^{22},~h^{12}=h^{21}$. The resulting linear (P-odd) response of the stress-energy tensor can be expressed as
\beq
&~&T_{11}=-T_{22}=i \xi_T(\omega, q)q_3~h_{12}\notag\\
&~&T_{12}=-i \xi_T(\omega, q)q_3~h_{11}\,.
\label{eq:simple_form}
\eeq
Here $\xi_T$ is proportional to the axial chemical potential $\mu$, which is used as a measure of the chiral asymmetry and is related to the P-odd part of the graviton self-energy. In what follows we refer to this transport as chiral gravitational effect (CGE).

While it is theoretically motivated to study the back-reaction of a chiral medium to a propagating gravitational perturbation, in general, one also could think about possible applications in the physics of early universe where
 chiral imbalance is often discussed in the context of axion dynamics and  primordial magnetic field generation \cite{Joyce:1997uy}. It is also appealing since all matter fields are coupled to the gravitational sector and one may expect P-odd effects due to the imbalance caused by a neutrino background appearing in some cosmological models, see e.g. \cite{Lesgourgues:1999wu, Maleknejad:2014wsa, MarchRussell:1999ig, McDonald:1999in}. 

Here we study the modification of the GW dispersion relation by a chiral medium at finite axial chemical potential $\mu$ and temperature $T$ concentrating on the P-odd features of the spectrum. We find that for time-like four momenta the dispersion relation modified by the P-odd part of the graviton self-energy results in a polarization dependence of GW damping. We compare this effect with an order-of-magnitude estimate of the P-even counterpart due to the medium viscosity \cite{Weinberg:2003ur, Baym:2017xvh} and find that the chiral damping of GWs is suppressed compared to the helicity-independent damping. Strikingly, for space-like four momenta, we find that the GW group velocity $v_g$ not only is helicity dependent but can turn negative for waves of particular polarization at a given sign of the chiral asymmetry. We stress that plasmons (electromagnetic modes) in chiral media exhibit the same peculiar behavior which is not discussed in the literature to the best of our knowledge\footnote{A detailed discussion of plasmons in chiral media can be found in \cite{Gorbar:2016sey, Gorbar:2016ygi}.}. The corresponding modification in the spectrum can be seen as a new helicity-dependent realization of the anomalous dispersion which attracted considerable attention in the context of artificial materials \cite{PhysRevA.63.053806, OptSocAmB}. It should be mentioned that propagation of helical waves in a chiral medium involves no gain or absorption in contrast with textbook examples of the anomalous dispersion \cite{Brillouin}. Finally, we argue that the helicity dependent birefringence may lead to specific phenomenological signatures which can in principle be observed. We stress that while in the case of gravity the effects are rather small due to the weakness of the gravitational interaction, the electromagnetic dispersion relation can be studied in tabletop experiments with topological systems of condensed matter supporting relativistic spectrum.

This paper is organized in the following way: in the next section we discuss an intuitive picture of the spin-gravity interaction leading to CGE. Then we turn to the details of the gravitational dispersion relation and show that the GW group velocity can turn negative. Finally, this result is generalized to the well-studied case of electromagnetic excitations in chiral media. We conclude our work with an outlook and a brief discussion of possible phenomenological consequences of the helicity dependent spectrum in chiral media.

\section{Chiral Matter in Gravitational Field} 
In order to understand the back-reaction of a chiral medium to a metric perturbation it is instructive to review the simpler case of the magnetic response. Considering a system of massive fermions in the presence of an external magnetic field $B$ one expects a net spin polarization. Indeed, the Pauli interaction $\delta E=-\CM\cdot B$, where $\CM$ is the magnetic moment, leads to a net spin alignment decreasing the energy of the system. Thus, at finite density it is natural to expect an average current of axial charge directly related to the spin polarization \cite{Metlitski:2005pr}. On the other hand, in a medium with a chiral imbalance one would find an electric current along the magnetic field which decays with time. 

Extending this argument, one would expect that CGE is sourced by a similar mechanism. Indeed, one may concentrate on the coupling of the stress-energy tensor of the medium to an external gravitational field $\mathcal L=-\frac{\kappa}{2}h^{\mu\nu}T_{\mu\nu}$. Following \cite{Berends:1975ah}, the expectation of the fermionic stress-energy tensor can be decomposed as
\beq
&&\langle P+q/2|T^{\mu\nu}(0)|P-q/2\rangle \nn\\
&&=-i\frac{\kappa}{4}\bar{u}(P+q/2)\bigg[A(q^2)\gamma^{\{\mu}P^{\nu\}}+\frac{2}{m}B(q^2)p^\mu p^\nu \nn\\
&& \,\,\,\,\,\,\,\,\,\,\,\,\,\,\,\,\,\,\,\,\,\,\,\, C(q^2)\frac{1}{m}(q^{\mu}q^{\nu}-g^{\mu\nu}q^2)\bigg]u(P-q/2)\nn\\
\label{ff}
\eeq 
where $q$ is momentum transfered to a fermion by the external field, $\{,\}$ denotes symmetrization of Lorentz indices and $\left(A,~B,~C\right)$ are gravitational form factors. The coupling of spin with a transverse-traceless (TT) metric perturbation\footnote{The interaction between spin and a GW was recently discussed in \cite{Obukhov:2017avp}} can be obtained from (\ref{ff}) in a fashion analogius to the derivation of the Pauli interaction from electromagnetic form factors. Taking the limit $q\rightarrow 0$ and using the non-relativistic expansion for the spinors one finds the energy due to the spin-gravity interaction which can be written as
\beq
\delta E \sim \epsilon^{\{ik3}\CM_g^{k}p^{j\}}q^3h^{ij}\,.
\label{Energy}
\eeq
From (\ref{Energy}) one expects a perturbation in $h^{11}$ to result in a preferred orthogonal combination of spin and linear momentum leading to non-zero perturbation in $\langle T^{12}\rangle$ according to (\ref{ff}). This effect cancels for two helicities of fermions in a P-even set-up but one may expect a response analogous to (\ref{eq:simple_form}) if there is a chiral imbalance.

Note that these simple arguments based on the spin-field interaction cannot serve as a rigorous derivation for massless fermions and should be supported by direct calculations in a chiral medium (as it is done for both magnetic \cite{Vilenkin:1980fu, Fukushima:2008xe, Metlitski:2005pr} and gravitational \cite{Manes:2012hf} responses). However, we think that this picture can be helpful for understanding of the origin of CGE.

\section{Gravitational Dispersion Relation}
Let us now analyse the spectrum of a TT metric perturbation propagating in a chiral plasma. The stress-energy tensor induced by the medium response (\ref{eq:simple_form}) modifies Einstein's equation for $h_{\rho\sigma}$. Following the standard Kubo formalism it can be expressed as $\langle T^{\mu\nu}\rangle =\Pi^{\mu\nu\rho\sigma}h_{\rho\sigma}$, where $\Pi^{\mu\nu\rho\sigma}$ is the retarded graviton self-energy. The P-odd TT component of the response function is derived in \cite{Manes:2012hf} and its tensorial structure reads
\beq
\label{TTpodd}
\Pi^{\mu\nu\rho\sigma}_T(\omega, q)=i \xi_T(\omega, q)u_\alpha Q_\beta\epsilon^{\alpha\beta\{\mu\rho}P_T^{\nu\}\sigma}+(\rho\leftrightarrow\sigma)\,,
\eeq
where $\xi_T$ is the scalar response function. One can see that (\ref{TTpodd}) explicitly satisfies the Ward identity which constraints the graviton self-energy and for the P-odd contribution is given by $q_\mu \Pi^{\mu\nu\rho\sigma}=0$.

Equipped with the response (\ref{TTpodd}), the linearized Einstein equation including back-reaction takes the form
\beq
G^{\text{pert}}_{\mu\nu}=\kappa T^{\text{pert}}_{\mu\nu}
\label{Eeq}
\eeq
where $\kappa$ is the gravitational coupling, $T^{pert}_{\mu\nu}$ and $G^{\text{pert}}_{\mu\nu}$ are the perturbations of the stress-energy tensor and the Einstein tensor. Note that in general $T^{\text{pert}}_{\mu\nu}$ contains also P-even contributions.

For a TT metric perturbation propagating along $z$-axis the linearised Einstein tensor takes a particularly simple form $G^{\text{pert}}_{ij}=(\omega^2-q^2)h_{ij}$ with $i(j)=1, 2$. The equation (\ref{Eeq}) is homogeneous with respect to $h^{\mu\nu}$ and has a non-trivial solution only if its determinant is zero. As usual, this constraint defines the spectrum of perturbations which is given by
\beq
\left(\omega^2-q^2+\frac{1}{m_p^2}\xi_S(\omega, q)\right)=\pm\frac{q}{m_p^2}\xi_T(\omega, q)\,
\label{disp}
\eeq
where the $\pm $ signs corresponds to two helicities, we express the gravitational coupling through the Planck mass $\kappa=m_p^{-2}$, and $\xi_S$ is introduced to stress the presence of P-even counterpart of the graviton self-energy. The leading contribution to $\xi_T$ in the derivative expansion $|\omega|^2, q^2\ll \mu^2, T^2$ was obtained in  \cite{Manes:2012hf} and is given by
\beq
\xi_T(\omega, q)=-\frac{1}{96\pi^2}\mu\left(\mu^2+\pi^2T^2\right)\left(2+\frac{Q^2}{q^2}+\frac{3Q^4}{q^4}L(\omega, q)\right)\notag\\
\label{cT}
\eeq
where
\beq
L(\omega, q)=-1+\frac{\omega}{2q}\log|\frac{\omega+q}{\omega-q}|-\frac{i\pi}{2}\frac{\omega}{q}\theta\left(1-\frac{\omega^2}{q^2}\right).
\eeq
with $Q^2=-\omega^2+q^2$.

Due to non-linearities in the spectrum it is natural to concentrate on physically motivated limits. Starting with the quasi-static regime $|\omega|\ll q$ we can write the leading contributions to the response function (\ref{cT}) as
\beq
\xi_T\simeq \frac{i\mu(\mu^2+\pi^2 T^2)}{64\pi}\frac{\omega}{q}\,.
\label{quasi}
\eeq
Note that in the strictly static limit of $\omega=0$ the P-odd part of the graviton self-energy $\xi_T$ goes to zero in contrast with the case of the photon self-energy responsible for CME. If P-even contributions are omitted from the consideration, the dispersion relation reduces to a simple form
\beq
\omega=\pm i q^2 \frac{64\pi m_p^2}{\mu(\mu^2+\pi^2 T^2)}\,.
\label{instb1}
\eeq
which indicates an exponentially growing mode similar to the chiral magnetic instability, see e.g. \cite{Akamatsu:2013pjd}. However, one would expect it to vanish if the P-even part of the back-reaction is taken into account. Then (\ref{instb1}) should be reinterpreted as a polarization dependent damping of modes in the quasi-static regime. It should be mentioned that Eq. \ref{instb1} implies $\frac{m_p^2 q}{\mu(\mu^2+\pi^2T^2)}\ll 1$ and characteristic wavelengths are expected to be larger than the horizon size $m_p/T^2$ even if the instability has a realization.

Note that in the strictly static limit $\xi_T$ is non-zero beyond the leading order in the derivative expansion \cite{Jensen:2012kj, Manes:2012hf}, the first non-trivial contribution appears at the second order and takes the form
\beq
\xi_T(0, q)=-\frac{1}{192\pi^2}\mu q^2\,.
\label{CGE3}
\eeq
One can readily find that for momenta satisfying $q^2-\frac{\mu q^2q_3}{192\pi^2m_p^2}<0$ there may appear a new type of chiral instability. To be specific, in the presence of the P-even damping, say due to viscosity, in the quasi-static limit, the spectrum could exhibit a negative imaginary part of the frequency, depending on the sign of the chiral imbalance leading to a growing mode. This instability is chiral in its nature and corresponds to an exponentially growing mode for a helicity fixed by the sign of $\mu$. However, its realization requires trans-Planckian momenta  $q>\frac{m_p^2}{\mu}$ in contrast with the chiral plasma instability leading beyond the applicability of the classical gravity and to a breakdown of the derivative expansion.

We now turn to the regime of propagating  waves $|\omega| \sim q$. Anticipating the imaginary part of the frequency $\omega_{\text{Im}}$ to be much smaller than the real part $\omega_{\text{Re}}$ in the expansion $Q^2/q^2$, the dispersion relation can be solved for the real part of the frequency with
\beq
\omega_{\text{Re}}^2 -q^2=\pm\frac{q}{m_P^2}\,\,\text{Re}[\xi_T(\omega_{\text{Re}}, q)].
\label{omegar}
\eeq
It is convenient to expand around the linear spectrum $\omega=q+\delta\omega$ with $\delta\omega_{\text{Re}/\text{Im}}$ being the real/imaginary part of the correction, then one finds
\beq
\delta\omega_{\text{Re}}=\pm\frac{1}{96\pi^2}\frac{\mu\left(\mu^2+\pi^2T^2\right)}{m_P^2}\,.
\label{REcorr}
\eeq
The two opposite chiralities result in distinct situations with $\omega_{\text{Re}}\lessgtr q$ depending on the sign of $\mu$. Indeed, neglecting the imaginary part of $\omega$, one expects that $\delta\omega_{\text{Re}}<q$ supports $\delta\omega_{\text{Im}}\neq 0$ while $\delta\omega_{\text{Re}}>q$ results in the theta function to be zero and $\delta\omega_{\text{Im}}= 0$. The leading contribution to the imaginary part of the frequency $\delta\omega_{\text{Im}}$ is given by
\beq
\delta\omega_{\text{Im}}=\frac{q}{2m_P^2}\frac{\text{Im}[\xi_T(q,\omega)]}{\omega_\text{Re}}\sim \frac{1}{q^2}\left(\frac{\mu\left(\mu^2+\pi^2T^2\right)}{m_P^2}\right)^3\,.
\label{Im}
\eeq
Although the dimensionful quantity multiplying the step function in the expression for $\xi_T$ can be positive or negative depending on the sign of the chemical potential $\mu$, the step function is non-zero only for one of the two polarizations (at given sign of $\mu$) fixing the sign of the imaginary part. Note that for positive helicity, a positive $\mu$ produces a negative $\delta\omega_{Re}$ and also sets the imaginary part of the frequency to be positive. This complex frequency corresponds to a suppression for one of polarized modes. The opposite helicity at $\mu>0$ produces a positive $\delta \omega_\text{Re}$ leading to $\delta\omega_{\text{Im}}=0$ due to the theta function in $L(\omega, q)$. In the presence of P-even contributions to the graviton self-energy, Eq. \ref{Im} leads to a helicity dependent damping of GWs similarly to the quasi-static limit.

While the helicity dependent damping seems to be suppressed by additional powers of $m_P$, it is instructive to roughly compare its magnitude with the effect of the P-even response in the same limit. One expects that in a P-even setup, both helicities of GWs are damped either by a Landau damping in the collision-less limit or by the viscous damping due to the shear viscosity $\eta$ of the medium, for additional details see the discussion in \cite{Baym:2017xvh}. For an estimate we concentrate on the effect of the shear viscosity $\eta$ or more precisely on the upper bound for the damping rate $\delta\omega_{\text{Im}}\sim \frac{\eta}{m_P^2}$ \cite{Hawking:1966qi}. For a rough estimate one can take $\eta\sim T^3$ and set $\mu\sim T$ reducing the number of scales in the problem. Then the helicity dependent contribution to the imaginary part of the frequency (\ref{Im}) is comparable with the viscous damping contribution only if
\beq
\frac{T^3}{m_P^2}\sim \frac{1}{q^2}\left(\frac{T^3}{m_P^2}\right)^3
\label{comp}
\eeq
or equivalently for $q\sim \frac{T^3}{m_P^2}$. In the limit $T\to m_P$ the relation (\ref{comp}) results in a characteristic momentum $q$ of the Planck scale. For modes of the horizon size $l\sim m_P/T^2$, the polarization dependent damping is suppressed comparing to the upper bound on the viscous damping by a small factor of $T^2/m_P^2$ decreasing for shorter wavelengths. 

In the long-wavelength regime $|\omega|/q\gg 1$ the leading GW dispersion relation is expected to be plasmon-like $\omega^2=\omega_{pl}^2+q^2$, see e.g. \cite{Rebhan:1990yr}. The plasma frequency can be estimated as $\omega_{pl}\sim \frac{T^4}{m_P^2}$ for $\mu\ll T$ and we focus on this limit. In the presence of a chiral asymmetry the spectrum is modified by a helicity-dependent P-odd contribution to the response function
\beq
\xi_T(q^0,q)=-\frac{1}{60}\mu T^2
\eeq
resulting in
\beq
\omega^2=\omega_{\text{pl}}^2 +q^2\pm \frac{1}{60}\frac{\mu T^2}{m_P^2}q\,.
\label{qdisp}
\eeq
Note that the linear term in powers of momentum $q$, on the RHS of (\ref{qdisp}), is much smaller than $\omega_{\text{pl}}^2$ for $\omega, q\ll \mu, T$ insuring stability of the system.

Propagation of long wavelength GWs through a chiral medium can be illustrated with a simple analysis of a wave-packet behavior. The presence of the linear term in the dispersion relation points to the dependence of group and phase velocities on the GW polarization. For simplicity we consider modes with positive circular polarization recovering results for the opposite helicity by the parity transformation (and changing the sign of $\mu$). Then the group velocity corresponding to the dispersion relation (\ref{qdisp}) is given by
\beq
v_g=\frac{d\omega}{dq}=\frac{q + \frac{1}{120}\frac{\mu T^2}{m_P^2}}{\sqrt{q^2+\omega_{\text{pl}}^2+\frac{1}{60}\frac{\mu T^2}{m_P^2}q}}
\label{group}\,.
\eeq
Remarkably, for $q< \frac{1}{120}\frac{\mu T^2}{m_P^2}$ the GW group velocity (\ref{group}) turns negative for $\mu<0$ indicating an anomalous dispersion for the positive polarization. One can see that the phase velocity $v_p\equiv\frac{\omega}{q}$ derived from (\ref{qdisp}) is also helicity dependent but the helical contribution is suppressed by $(\mu q)/T^2$ comparing to the leading P-even term.

Before we delve into the implications of the chiral anomalous dispersion, let us estimate the order of magnitude of the terms involved in (\ref{qdisp}) in the regime where group velocity turns negative. The corresponding momentum is given by $q\sim \mu T^2/m_p^2$ and satisfies both $q\ll (\mu, T)$ and $q\ll \omega_{pl}$ for $(\mu, T)\ll m_P$. In this limit one finds that $\omega \sim \omega_{pl}$ and the expansion parameter $q/\omega \sim q/\omega_{pl}\sim \mu/m_p$ is small as expected. 

It is well known that Gaussian packets of propagating waves travel with the group velocity which, if negative, can lead to a peculiar behaviour. The scalar part of the Fourier amplitude for a gaussian wave-packet centered around the momentum $q_c$ is given by $e^{-\alpha(q-q_c)^2}$. An inverse Fourier-transform produces the following amplitude in coordinate space
$$e^{i(q_c z-\omega_0 t)}e^{-\frac{(z-v_g(q_c) t)^2}{4\alpha}}$$ 
after expanding the spectrum around $q=q_c$. Thus, if the group velocity changes sign, the wave-packets for two polarizations propagate in the opposite directions.

Typically a negative group velocity is associated with an absorptive medium. The simplest example \cite{Brillouin} involves modeling the response of atoms or molecules in ordinary matter to an external perturbation of electromagnetic field. The atoms or molecules gain dipole moments and are treated as damped simple harmonic oscillators with some characteristic absorption frequency. In this case the group velocity and the refractive index $n$ are complex for propagating modes with a real wave-vector.  Near the characteristic frequency, the real part of the index of refraction jumps from a positive to a negative value. The slope of the real part of the refractive index becomes negative with increasing frequency, causing the group velocity to become negative and at times infinite $v_g=(n+\omega \frac{dn}{d\omega})^{-1}$. However, the chiral anomalous dispersion for GWs obtained in this paper is not associate with any absorption. The frequency, the index of refraction and the group velocity remain real.

Although a negative group velocity in chiral medium will require $q<\frac{1}{120}\frac{\mu T^2}{m_p^2}$, chiral splitting of group velocities is present for arbitrary $q<\omega$. If $q\gtrsim\frac{1}{120}\frac{\mu T^2}{m_p^2}$ the group velocity is positive for both helicities but is still polarization dependent. In order to estimate the upper bound on the helical correction to $v_g$ one can set $q\sim\omega_{pl}\sim T^2/m_p$. Then the chiral contribution to the group velocity goes as $\mu/m_P$ while the leading term is of order $\omega_{pl}/q\sim 1$. An accurate estimate of the chiral splitting of the group velocity requires numerical study of the spectrum including the full self-energy and we leave it for future work. 

\section{Electromagnetic Dispersion Relation}
Here we briefly discuss propagation of electromagnetic modes in a chiral plasma in the presence of non-zero $\mu$ and show that the chiral anomalous dispersion is a general feature of helical excitations in such systems. The dispersion relation for plasmons can be obtained in a manner similar to the GW case discussed in the previous section. Maxwell's equations with the back-reaction taken into account produce
\beq
\text{det}\left[(q^2-\omega^2)\delta^{ij}-k^i k^j+\Pi^{ij}\right]=0\,,
\label{eq:pdisp}
\eeq
and we use $A^0=0$ gauge following conventions of \cite{Akamatsu:2013pjd}. Here the response function $\Pi^{ij}$ is the retarded photon self-energy in a chiral medium derived in \cite{Son:2012zy}. At finite temperature and density there is a preferred reference frame, the polarization operator has a more involved structure than in vacuum and can be decomposed into longitudinal, transverse, and antisymmetric parts
\beq
&&\Pi^{ij}=\Pi_L P_Li^{ij}+\Pi_T P_Ti^{ij}+\Pi_A P_Ai^{ij}\notag\\
&&P_L^{ij}=\frac{q^i q^j}{q^2}~,~P_T=\delta^{ij}-P_L^{ij}~,~P_A^{ij}=\frac{i\epsilon^{ijk}q_k}{q}\,,
\eeq
where the P-odd contribution corresponds to the presence of a chiral imbalance. Using this decomposition one can reduce (\ref{eq:pdisp}) to
\beq
\omega^2=\Pi_L~~,~\omega^2=q^2+\Pi_T\pm\Pi_A
\label{eq:spec}
\eeq
with two signs corresponding to different polarizations. The explicit form the three contributions to the photon self-energy are given by
\beq
\Pi_L=m_D^2\frac{\omega^2}{q^2}\tilde{L}(\omega,q)~,~~\Pi_T=\frac{m_D^2}{2}\left(1+\frac{q^2-\omega^2}{q^2}\tilde{L}(\omega,q)\right)~,~~\Pi_A=-\frac{\alpha \mu q}{\pi}\left(1-\frac{\omega^2}{q^2}\right)\tilde{L}(\omega,q)\,,\notag\\
\eeq
where $m_D^2=e^2\left(\frac{T^2}{6}+\frac{\mu^2}{2\pi^2}\right)$, $\tilde{L}(\omega,q)\equiv L(\omega,q)+1$ and we take the wave-vector in the $z$-direction. In analogy with the gravity sector, we consider the photon self-energy in the long-wavelength limit $|\omega/q|\gg 1$,
\beq
\Pi_L\simeq \frac{1}{3}m_D^2+\CO\left(\frac{q^2}{\omega^2}\right)~,~~\Pi_T\simeq \frac{1}{3}m_D^2+\CO\left(\frac{q^2}{\omega^2}\right)~,~~\Pi_A\simeq \frac{\alpha \mu q}{3\pi}+\CO\left(\frac{q^3}{\omega^3}\right)\,.
\label{eq:Pexp}
\eeq
Substituting (\ref{eq:Pexp}) into (\ref{eq:spec}) one finds the dispersion relation in the long-wavelength limit
\beq
\omega^2=\omega_{\text{pl}}^2+q^2\pm\frac{\alpha \mu q}{3\pi}+\mathcal{O}\left(\frac{q^2}{\omega^2}\right)\,.
\label{eq:pspec}
\eeq
As previously in the GW case, the dispersion relation is helicity dependent and one of the two polarizations acquires a negative group velocity for $q<\frac{\alpha}{3\pi}\mu$. Thus, we have shown that the chiral anomalous dispersion discussed in this papers exists for the electromagnetic modes in chiral media as well as for gravitational ones. Emphasizing a point mentioned earlier, a milder constraint $\omega > q$ is satisfied for $q\sim\omega_{pl}$. Although this regime is not accessible analytically, a numerical analysis is expected to reproduce the chiral splitting of the group velocity and we leave it for further study. The relatively larger value of the electromagnetic coupling makes the anomalous dispersion of plasmons in chiral media to be in principle observable in tabletop experiments based on condensed matter systems with relativistic linear spectrum such as Dirac semi-metals. This proposal is additionally supported by the recent experimental observation of CME in those systems, see \cite{Li:2014bha}.

\section{Outlook and Discussion}

In this paper we study how circularly polarized electromagnetic and gravitational waves propagate through a chiral medium. The response of a chiral medium to the electromagnetic and gravitational fields is given by photon and graviton retarded self-energies and we use that to construct linearized equations of motion for corresponding waves. These equations determine the spectrum of excitations propagating in a chiral plasma.

We start with the gravitational sector and consider CGE -- the chiral medium response to an external TT gravitational field in the stress-energy tensor. We consider the back reaction of this effect to gravitational perturbations propagating through the medium and the resulting spectrum in several limiting regimes. Concentrating on P-odd features of the dynamics we mostly omit the P-even part of the graviton self-energy where it is possible and qualitatively restore it if required. In the quasi-static limit the electromagnetic response results in the chiral magnetic instability which corresponds to a transfer of the microscopic chirality to the macroscopic helicity of magnetic field, see e.g. \cite{Akamatsu:2013pjd}. Following this intuition we start the consideration of GWs in a chiral medium with the quasi-static limit $|\omega|/q\ll 1$ and find that a gravitational chiral instability would require wavelength above the horizon size. We further argue that this regime is also considerably modified by P-even contributions and the unstable behavior vanishes. This is expected since the leading order CGE disappears in the exact static limit in contrast with CME. The first non-trivial CGE contribution in the static limit appears at the 3d order in the derivative expansion (\ref{CGE3}). Taking it into account one can find that the modified dispersion relation exhibits another instability (in the UV limit) which, however, requires trans-Planckian momenta and goes beyond applicability of the classical gravity (and this consideration). In the regime of relativistic propagating waves $\omega\sim q$ the spectrum indicates helicity dependent attenuation which leads to an unequal damping of polarized GWs. It is however highly suppressed for shorter wavelengths. We illustrate this result comparing the helicity dependent contribution with the textbook example of the GW damping by the shear viscosity. Maximizing the P-odd effect we consider modes of the horizon size and find that it is still suppressed by  a small factor of $T^2/m_P^2$ compared to the P-even part. Turning to the long-wavelength limit $|\omega|/q\gg 1$ we find an interesting result -- the GW group velocity appears to be helicity dependent and may become negative giving rise to a new type of anomalous dispersion. We believe that the mutual effect of the polarization dependent GW propagation can turn into an overall helical asymmetry in distinct regions of space which in principle could lead to helical distributions of matter and radiation. One should expect this asymmetry to be rather small but it would be interesting to study whether its signatures could at least in principle be observed. We also note that an additional chiral imbalance may be caused by a background electromagnetic helicity\footnote{For a recent discussion of chiral effects for photons and an electromagnetic helicity generation, see \cite{Avkhadiev:2017fxj, Yamamoto:2017uul}.} and axions which should be added to this consideration.

Inspired by the group velocity behavior in the gravitational case we turn to reviewing electromagnetic excitations in chiral media. The plasmon spectrum in a medium with a chiral asymmetry is studied in the literature in great details, for a recent discussion see \cite{Gorbar:2016sey, Gorbar:2016ygi}. Focusing on the long-wavelength limit we find that the leading correction to the plasmon spectrum, indeed, results in a similar phenomenon: the plasmon group velocity of a particular helicity can turn negative. The electromagnetic coupling is much larger resulting in a stronger effect than that in the gravitational sector and one may expect that this birefringence of chiral media may be observed in experiments with Dirac semi-metals. In addition, phenomenological effects of the chiral anomalous dispersion can be studied for a radiation produced in and passing through a cosmic P-odd background such as a supernova where a chiral asymmetry may be generated. Then, one can study the difference between polarized light signals searching for helical asymmetries.

\section{Acknowledgments}
We would like to thank Yi Yin for initializing the discussion on the GW propagation in chiral media. We would like to thank D.B. Kaplan, V. Kirilin, L. McLerran, S. Reddy, I. Shovkovy, O. Teryaev, and V. I. Zakharov for useful discussions and comments. AS is supported by the LANL/LDRD Program and, at the beginning of this work, by U.S. Department of Energy under grant Contract Number DE-SC0011090. AS is grateful for travel support from RFBR grant 17-02-01108 at the beginning of this work. SS is supported by U.S. Department of Energy under grant Contract Number DE-FG02-00ER41132 and, at the beginning of this work, by U.S. Department of Energy under grant Contract Number DE-FG02-04ER41338.

\newpage

\bibliographystyle{bibstyle}
\bibliography{CGW}

\begin{thebibliography}{10}
\ifx\href\asklfhas\newcommand{\href}[2]{#2}\fi
\ifx\arxivref\asklfhas\newcommand{\arxivref}[2]{\href{http://arxiv.org/abs/#1}{#2}}\fi
\ifx\doiref\asklfhas\newcommand{\doiref}[2]{\href{http://dx.doi.org/#1}{#2}}\fi
\parskip 0pt
\normalsize

\bibitem{Kharzeev:2015znc}
D.~E. Kharzeev, J.~Liao, S.~A. Voloshin \& G.~Wang,
\textit{``{Chiral magnetic and vortical effects in high-energy nuclear
  collisions -- A status report}''},
\doiref{10.1016/j.ppnp.2016.01.001}{Prog.~Part.~Nucl.~Phys. \textbf{88}, 1
  (2016)},
\normalsize{\texttt{\arxivref{1511.04050}{arXiv:1511.04050}}}.

\bibitem{Joyce:1997uy}
M.~Joyce \& M.~E. Shaposhnikov,
\textit{``{Primordial magnetic fields, right-handed electrons, and the Abelian
  anomaly}''},
\doiref{10.1103/PhysRevLett.79.1193}{Phys.~Rev.~Lett. \textbf{79}, 1193
  (1997)},
\normalsize{\texttt{\arxivref{astro-ph/9703005}{astro-ph/9703005}}}.

\bibitem{Li:2014bha}
Q.~Li, D.~E. Kharzeev, C.~Zhang, Y.~Huang, I.~Pletikosic, A.~V. Fedorov, R.~D.
  Zhong, J.~A. Schneeloch, G.~D. Gu \& T.~Valla,
\textit{``{Observation of the chiral magnetic effect in ZrTe5}''},
\doiref{10.1038/nphys3648}{Nature~Phys. \textbf{12}, 550 (2016)},
\normalsize{\texttt{\arxivref{1412.6543}{arXiv:1412.6543}}}.

\bibitem{Akamatsu:2013pjd}
Y.~Akamatsu \& N.~Yamamoto,
\textit{``{Chiral Plasma Instabilities}''},
\doiref{10.1103/PhysRevLett.111.052002}{Phys.~Rev.~Lett. \textbf{111}, 052002
  (2013)},
\normalsize{\texttt{\arxivref{1302.2125}{arXiv:1302.2125}}}.

\bibitem{Khaidukov:2013sja}
Z.~V. Khaidukov, V.~P. Kirilin, A.~V. Sadofyev \& V.~I. Zakharov,
\textit{``{On Magnetostatics of Chiral Media}''},
\normalsize{\texttt{\arxivref{1307.0138}{arXiv:1307.0138}}}.

\bibitem{Kirilin:2013fqa}
V.~P. Kirilin, A.~V. Sadofyev \& V.~I. Zakharov,
\textit{``{Anomaly and long-range forces}''},
in \textit{``{Proceedings, 100th anniversary of the birth of I.Ya. Pomeranchuk
  (Pomeranchuk 100): Moscow, Russia, June 5-6, 2013}''},
p.~272-286.
\bibitem{Tuchin:2014iua}
K.~Tuchin,
\textit{``{Electromagnetic field and the chiral magnetic effect in the
  quark-gluon plasma}''},
\doiref{10.1103/PhysRevC.91.064902}{Phys.~Rev. \textbf{C91}, 064902 (2015)},
\normalsize{\texttt{\arxivref{1411.1363}{arXiv:1411.1363}}}.

\bibitem{Avdoshkin:2014gpa}
A.~Avdoshkin, V.~P. Kirilin, A.~V. Sadofyev \& V.~I. Zakharov,
\textit{``{On consistency of hydrodynamic approximation for chiral media}''},
\doiref{10.1016/j.physletb.2016.01.048}{Phys.~Lett. \textbf{B755}, 1 (2016)},
\normalsize{\texttt{\arxivref{1402.3587}{arXiv:1402.3587}}}.

\bibitem{Buividovich:2015jfa}
P.~V. Buividovich \& M.~V. Ulybyshev,
\textit{``{Numerical study of chiral plasma instability within the classical
  statistical field theory approach}''},
\doiref{10.1103/PhysRevD.94.025009}{Phys.~Rev. \textbf{D94}, 025009 (2016)},
\normalsize{\texttt{\arxivref{1509.02076}{arXiv:1509.02076}}}.

\bibitem{Kaplan:2016drz}
D.~B. Kaplan, S.~Reddy \& S.~Sen,
\textit{``{Energy Conservation and the Chiral Magnetic Effect}''},
\doiref{10.1103/PhysRevD.96.016008}{Phys.~Rev. \textbf{D96}, 016008 (2017)},
\normalsize{\texttt{\arxivref{1612.00032}{arXiv:1612.00032}}}.

\bibitem{Hattori:2017usa}
K.~Hattori, Y.~Hirono, H.-U. Yee \& Y.~Yin,
\textit{``{Magnetoydroynamics with chiral anomaly: phases of collective
  excitations and instabilities}''},
\normalsize{\texttt{\arxivref{1711.08450}{arXiv:1711.08450}}}.

\bibitem{Yamamoto:2015gzz}
N.~Yamamoto,
\textit{``{Chiral transport of neutrinos in supernovae: Neutrino-induced fluid
  helicity and helical plasma instability}''},
\doiref{10.1103/PhysRevD.93.065017}{Phys.~Rev. \textbf{D93}, 065017 (2016)},
\normalsize{\texttt{\arxivref{1511.00933}{arXiv:1511.00933}}}.

\bibitem{Zakharov:2016lhp}
V.~I. Zakharov,
\textit{``{Notes on conservation laws in chiral hydrodynamics}''},
\normalsize{\texttt{\arxivref{1611.09113}{arXiv:1611.09113}}}.

\bibitem{Kirilin:2017tdh}
V.~P. Kirilin \& A.~V. Sadofyev,
\textit{``{Anomalous Transport and Generalized Axial Charge}''},
\doiref{10.1103/PhysRevD.96.016019}{Phys.~Rev. \textbf{D96}, 016019 (2017)},
\normalsize{\texttt{\arxivref{1703.02483}{arXiv:1703.02483}}}.

\bibitem{Basar:2013qia}
G.~Basar, D.~E. Kharzeev \& I.~Zahed,
\textit{``{Chiral and Gravitational Anomalies on Fermi Surfaces}''},
\doiref{10.1103/PhysRevLett.111.161601}{Phys.~Rev.~Lett. \textbf{111}, 161601
  (2013)},
\normalsize{\texttt{\arxivref{1307.2234}{arXiv:1307.2234}}}.

\bibitem{Flachi:2017vlp}
A.~Flachi \& K.~Fukushima,
\textit{``{Chiral vortical effect in curved space and the Chern-Simons
  current}''},
\normalsize{\texttt{\arxivref{1702.04753}{arXiv:1702.04753}}}.

\bibitem{Landsteiner:2011cp}
K.~Landsteiner, E.~Megias \& F.~Pena-Benitez,
\textit{``{Gravitational Anomaly and Transport}''},
\doiref{10.1103/PhysRevLett.107.021601}{Phys.~Rev.~Lett. \textbf{107}, 021601
  (2011)},
\normalsize{\texttt{\arxivref{1103.5006}{arXiv:1103.5006}}}.

\bibitem{Jensen:2012kj}
K.~Jensen, R.~Loganayagam \& A.~Yarom,
\textit{``{Thermodynamics, gravitational anomalies and cones}''},
\doiref{10.1007/JHEP02(2013)088}{JHEP \textbf{1302}, 088 (2013)},
\normalsize{\texttt{\arxivref{1207.5824}{arXiv:1207.5824}}}.

\bibitem{Manes:2012hf}
J.~L. Manes \& M.~Valle,
\textit{``{Parity violating gravitational response and anomalous constitutive
  relations}''},
\doiref{10.1007/JHEP01(2013)008}{JHEP \textbf{1301}, 008 (2013)},
\normalsize{\texttt{\arxivref{1211.0876}{arXiv:1211.0876}}}.

\bibitem{Lesgourgues:1999wu}
J.~Lesgourgues \& S.~Pastor,
\textit{``{Cosmological implications of a relic neutrino asymmetry}''},
\doiref{10.1103/PhysRevD.60.103521}{Phys.~Rev. \textbf{D60}, 103521 (1999)},
\normalsize{\texttt{\arxivref{hep-ph/9904411}{hep-ph/9904411}}}.

\bibitem{Maleknejad:2014wsa}
A.~Maleknejad,
\textit{``{Chiral Gravity Waves and Leptogenesis in Inflationary Models with
  non-Abelian Gauge Fields}''},
\doiref{10.1103/PhysRevD.90.023542}{Phys.~Rev. \textbf{D90}, 023542 (2014)},
\normalsize{\texttt{\arxivref{1401.7628}{arXiv:1401.7628}}}.

\bibitem{MarchRussell:1999ig}
J.~March-Russell, H.~Murayama \& A.~Riotto,
\textit{``{The Small observed baryon asymmetry from a large lepton
  asymmetry}''},
\doiref{10.1088/1126-6708/1999/11/015}{JHEP \textbf{9911}, 015 (1999)},
\normalsize{\texttt{\arxivref{hep-ph/9908396}{hep-ph/9908396}}}.

\bibitem{McDonald:1999in}
J.~McDonald,
\textit{``{Naturally large cosmological neutrino asymmetries in the MSSM}''},
\doiref{10.1103/PhysRevLett.84.4798}{Phys.~Rev.~Lett. \textbf{84}, 4798
  (2000)},
\normalsize{\texttt{\arxivref{hep-ph/9908300}{hep-ph/9908300}}}.

\bibitem{Weinberg:2003ur}
S.~Weinberg,
\textit{``{Damping of tensor modes in cosmology}''},
\doiref{10.1103/PhysRevD.69.023503}{Phys.~Rev. \textbf{D69}, 023503 (2004)},
\normalsize{\texttt{\arxivref{astro-ph/0306304}{astro-ph/0306304}}}.

\bibitem{Baym:2017xvh}
G.~Baym, S.~P. Patil \& C.~J. Pethick,
\textit{``{Damping of gravitational waves by matter}''},
\doiref{10.1103/PhysRevD.96.084033}{Phys.~Rev. \textbf{D96}, 084033 (2017)},
\normalsize{\texttt{\arxivref{1707.05192}{arXiv:1707.05192}}}.

\bibitem{Gorbar:2016sey}
E.~V. Gorbar, V.~A. Miransky, I.~A. Shovkovy \& P.~O. Sukhachov,
\textit{``{Chiral magnetic plasmons in anomalous relativistic matter}''},
\doiref{10.1103/PhysRevB.95.115202}{Phys.~Rev. \textbf{B95}, 115202 (2017)},
\normalsize{\texttt{\arxivref{1611.05470}{arXiv:1611.05470}}}.

\bibitem{Gorbar:2016ygi}
E.~V. Gorbar, V.~A. Miransky, I.~A. Shovkovy \& P.~O. Sukhachov,
\textit{``{Consistent Chiral Kinetic Theory in Weyl Materials: Chiral Magnetic
  Plasmons}''},
\doiref{10.1103/PhysRevLett.118.127601}{Phys.~Rev.~Lett. \textbf{118}, 127601
  (2017)},
\normalsize{\texttt{\arxivref{1610.07625}{arXiv:1610.07625}}}.

\bibitem{PhysRevA.63.053806}
A.~Dogariu, A.~Kuzmich \& L.~J. Wang,
\textit{``Transparent anomalous dispersion and superluminal light-pulse
  propagation at a negative group velocity''},
\doiref{10.1103/PhysRevA.63.053806}{Phys.~Rev.~A \textbf{63}, 053806 (2001)},
\href{https://link.aps.org/doi/10.1103/PhysRevA.63.053806}{\texttt{https://link.aps.org/doi/10.1103/PhysRevA.63.053806}}.

\bibitem{OptSocAmB}
W.~Brown, R.~McLean, A.~Sidorov, P.~Hannaford \& A.~Akulshin,
\textit{``Anomalous dispersion and negative group velocity in a coherence-free
  cold atomic medium''},
\doiref{10.1364/JOSAB.25.000C82}{Opt.~Soc.~Am.~B \textbf{25},  (2008)}.

\bibitem{Brillouin}
L.~Brillouin,
\textit{``Wave propagation and group velocity''}.

\bibitem{Metlitski:2005pr}
M.~A. Metlitski \& A.~R. Zhitnitsky,
\textit{``{Anomalous axion interactions and topological currents in dense
  matter}''},
\doiref{10.1103/PhysRevD.72.045011}{Phys.~Rev. \textbf{D72}, 045011 (2005)},
\normalsize{\texttt{\arxivref{hep-ph/0505072}{hep-ph/0505072}}}.

\bibitem{Berends:1975ah}
F.~A. Berends \& R.~Gastmans,
\textit{``{Quantum Electrodynamical Corrections to Graviton-Matter
  Vertices}''},
\doiref{10.1016/0003-4916(76)90245-1}{Annals~Phys. \textbf{98}, 225 (1976)}.

\bibitem{Obukhov:2017avp}
Y.~N. Obukhov, A.~J. Silenko \& O.~V. Teryaev,
\textit{``{General treatment of quantum and classical spinning particles in
  external fields}''},
\doiref{10.1103/PhysRevD.96.105005}{Phys.~Rev. \textbf{D96}, 105005 (2017)},
\normalsize{\texttt{\arxivref{1708.05601}{arXiv:1708.05601}}}.

\bibitem{Vilenkin:1980fu}
A.~Vilenkin,
\textit{``{EQUILIBRIUM PARITY VIOLATING CURRENT IN A MAGNETIC FIELD}''},
\doiref{10.1103/PhysRevD.22.3080}{Phys.~Rev. \textbf{D22}, 3080 (1980)}.

\bibitem{Fukushima:2008xe}
K.~Fukushima, D.~E. Kharzeev \& H.~J. Warringa,
\textit{``The Chiral Magnetic Effect''},
\doiref{10.1103/PhysRevD.78.074033}{Phys.~Rev.~D \textbf{78}, 074033 (2008)},
\normalsize{\texttt{\arxivref{0808.3382}{arXiv:0808.3382}}}.

\bibitem{Hawking:1966qi}
S.~W. Hawking,
\textit{``{Perturbations of an expanding universe}''},
\doiref{10.1086/148793}{Astrophys.~J. \textbf{145}, 544 (1966)}.

\bibitem{Rebhan:1990yr}
A.~Rebhan,
\textit{``{Collective phenomena and instabilities of perturbative quantum
  gravity at nonzero temperature}''},
\doiref{10.1016/S0550-3213(05)80041-0}{Nucl.~Phys. \textbf{B351}, 706 (1991)}.

\bibitem{Son:2012zy}
D.~T. Son \& N.~Yamamoto,
\textit{``{Kinetic theory with Berry curvature from quantum field theories}''},
\doiref{10.1103/PhysRevD.87.085016}{Phys.~Rev. \textbf{D87}, 085016 (2013)},
\normalsize{\texttt{\arxivref{1210.8158}{arXiv:1210.8158}}}.

\bibitem{Avkhadiev:2017fxj}
A.~Avkhadiev \& A.~V. Sadofyev,
\textit{``{Chiral Vortical Effect for Bosons}''},
\normalsize{\texttt{\arxivref{1702.07340}{arXiv:1702.07340}}}.

\bibitem{Yamamoto:2017uul}
N.~Yamamoto,
\textit{``{Photonic chiral vortical effect}''},
\doiref{10.1103/PhysRevD.96.051902}{Phys.~Rev. \textbf{D96}, 051902 (2017)},
\normalsize{\texttt{\arxivref{1702.08886}{arXiv:1702.08886}}}.

\end{thebibliography}

\end{document}